\documentclass[useAMS,usenatbib,usenatbib,usegraphicx]{mn2e}

\title[Dust Reddening in SDSS Spectra with Intervening DLA]{Searching for Dust Reddening in SDSS Spectra with Damped Lyman $\alpha${} Systems}
\author[S. Frank and C. P\'eroux]{S. Frank$^{1}$\thanks{E-mail: sfrank@oamp.fr} and C. P\'eroux$^{1}$\footnotemark[1]\\
$^{1}$Laboratoire d'Astrophysique de Marseille, OAMP, Universit\'e Aix-Marseille \&{} CNRS, 38 rue Fr\'ed\'eric Joliot Curie,\\
 13388 Marseille cedex 13, France}

\begin{document} 

\date{Accepted 2010 April 13. Received 2010 March 18; in original form 2009 December 07}

\pagerange{\pageref{firstpage}--\pageref{lastpage}} 

\maketitle

\label{firstpage}

\begin{abstract}
We searched for evidence of the reddening of background QSO spectra from the Sloan Digital Sky Survey (SDSS) due to dust in intervening damped Lyman $\alpha$ (DLA) systems. We utilise the two Data Releases DR5 and DR7 to arrive at sample sizes of 475 (DR5) and 676 (DR7) absorbers, based on two different published lists of SDSS DLAs. Both samples span roughly the redshift range of $2.2<z_{abs}<5.2$, with a mean of z$\sim$3.0, and the majority of the DLAs (75\%) below z=3.3. We construct geometric mean spectra in the absorber restframes ranging from 1240 to $\sim$ 2800 \AA, and composite spectra of samples matching the 'DLA' QSOs in $i${} band magnitude and emission redshift z$_{em}$, but without absorption lines. By comparing the slopes of these composite spectra with their matched counterparts, we find no sign of reddening in the ensemble of the absorbers from these samples. Owing to both the unprecedently large sizes of the DLA samples themselves and the non-DLA SDSS QSO sample, from which we can draw our matching spectra, we can place very tight limits for this non-detection ($<$E(B-V)$>${} =-0.0013$\pm$ 0.0025 (DR5) and $<$E(B-V)$>${} =-0.0017$\pm$0.0022 (DR7). Interestingly, when applying our technique to the samples of York et. al. (2006), vandenBerk et al. (2008) (intervening and intrinsic MgII absorbers) and the smaller DLA-subsample and pool of comparison QSOs of Vladilo et al. (2008), we do recover their results, i.e. detect the same amount of reddening as these authors do. Furthermore, we have tested whether subsamples of our large sample in categories involving the absorbers (HI column densities, presence or absence of accompanying metal absorption, absorber redshift) or the background quasars (emission redshift, brightness) do reveal dust extinction, but found no trends. These results are at odds with both detections of dust reddening  from previous studies, and also with expectations from observations of high-redshift galaxies.             
\end{abstract}

\begin{keywords}
Quasar absorption lines - Damped Lyman alpha systems - Dust
\end{keywords}

\section{Introduction}\label{introduction}
           
Detecting galaxies at high-redshift provides a powerful observational
way to probe structure formation in the early Universe. There have now
been several hundreds of galaxies found using the Lyman-break
technique (e.g. \citet[]{steidel1999}. Although very successful, this
method is limited to luminous objects. A complementary technique to
select high-redshift galaxies is the detection of systems in
absorption in the spectra of background quasars. Since their selection
is independent of luminosity and morphology, the study of galaxy
evolution through these systems circumvents the biases inherent to
traditional emission studies.\\

 Damped Lyman-$\alpha$ systems (hereafter
DLAs) have hydrogen column densities, log N(HI) $>$ 20.3 and are
thought to be associated with parts of high-redshift galaxies, just as they
are at lowest redshift \citep[]{rao2003, chen2005}. From our
knowledge of the local Universe we expect that high-redshift galaxies,
especially those forming stars, should contain some amount of
dust. Therefore, the possibility that DLAs are dusty objects is
currently hotly debated. Indeed, smoothed-particle hydrodynamic
simulations \citep[]{nagamine2004}{} predict that DLAs
should be 1/3$^{\rm rd}$ solar at $z=2.5$ and even more metal-rich
toward lower redshifts whilst current observations measure typically a
metallicity around 1/10$^{\rm th}$ solar. Dust is therefore invoked to
reconcile results from the simulations with current observations. In
addition, one can estimate the dust content of individual DLA from the
ratio of depleted onto undepleted elements it contains. It is
assumed that a fraction of the elements is not detected in the gas
phase because it is locked into dust grains (e.g. \citet[]{pettini2000}.
 Such studies have shown that the dust depletion levels
observed in DLAs are small, lower than those measured in the Galactic
interstellar-medium clouds with similar N(HI) column densities. In
fact, the observations resemble the dust depletion of the Galactic
halo warm gas \citep[]{kulkarni1997, savaglio2000}.

Another more direct way to probe the dust content of DLAs is to look
for a 2175 \AA\ feature (the so-called ``dust bump'') in the quasar
spectrum. This feature is known to be more predominant in the Milky
Way than in the Small and Large Magellanic Clouds. Many searches lead
to no-detections \citep[]{jura1977, smith1979, boisse1988, lanzetta1989, fall1989, pei1991}. Using Sloan Digital Sky Survey (SDSS) spectra, \citet[]{wang2004} claimed the detections of bumps in three individual MgII
absorbers at z=1.5, while \citet[]{junkkarinen2004}{} have detected a
clear bump in a DLA at z=0.524 toward a BL Lac at z=0.94. Other
searches remained unsuccessful, indicating that most DLAs do not show a
predominant 2175 \AA\ feature. A promising tool for finding the signatures of dust are searches for the
strong silicate absorption features in the infrared. \citet[]{kulkarni2007}, using
Spitzer data, report on the first indication of 9.7 $\mu$m silicate absorption in a DLA at z$_{abs}=0.52$, and find it to be
stronger than expected from extrapolation of the relation from Galactic diffuse ISM. 
\\

A consequence of DLAs being dusty is that it might prevent the
inclusion of their background quasar in colour-colour selected
samples. In order to overcome such biases, radio-selected quasar
samples have been built recently: while \citet[]{ellison2001}{} did not
discover a missing population, a more recent survey finds 4 fields
(out of $\sim$60) where the radio-flux selected quasars have no
optical counterparts down to very faint limits \citep[]{jorgenson2006}. These objects are the potential candidates of dusty DLAs
obscuring background quasars.\\  

In this paper, we investigate yet another possible signature of the
dust content of DLAs. Indeed, dust in a quasar absorber will redden
the colour of its background quasar. \citet[]{pei1991}{} were
the first to look for reddening of the quasar continua in a sample
containing DLAs and detect reddening at 99.999\% level. More recently,
\citet[]{murphy2004}{} have used 79 DLAs from SDSS Data Release 2 to
compare the slope of power laws fitting quasar continua with and
without DLAs. These authors found a colour excess due to Small
Magellanic Cloud type of dust of reddening $<$E(B-V)$>${} $<$0.01 mag. Using colour
excess measurements for SDSS sightlines containing DLAs and matching sightlines without them, \citet[]{vladilo2008} report a detection of reddening in a sample of DR5 SDSS QSOs towards sightlines containing DLAs at the mean level of $<$E(B-V)$>${} $\sim$6.3x10$^{-3}$. They derive their estimate by comparing colours of QSOs with intervening DLAs to the mean colour of a set of QSO of similar emission redshift and brightness, and bootstrapping the resulting broad distribution of colour excesses to ascertain the slightly redder colours of the DLA QSOs. A byproduct of the \citet[]{pontzen2009}{} work, who performed a Bayesian analysis on a variety of optical and radio datasets to determine the loss fraction of QSOs due to dust obscuration, is an estimate of the probability density distribution for dust reddening in optical spectra, that also peaks around the same value for $<$E(B-V)$>${} . Here, we
build composite spectra of quasar spectra with DLAs from
SDSS Data Releases 5 and 7, and compare them with reference composite spectra
built to match the original spectra in magnitudes and redshifts.

The organisation of the paper is as follows : after the introduction, we describe the selection of the DLA samples in section 2, followed by the construction of the matching non-absorber samples, and the procedure for obtaining composite spectra in section 3, where we also presents the method for deriving extinction as well as the results for the full samples. After discussing subsample analyses (section 4), we summarise and conclude in section 5.\\


\section{DLA sample selection}\label{dla_sample}
We start the selection of our samples from two lists of SDSS surveys for DLAs.

 \citet[]{prochaska2009} searched in Data Release (DR) 5, following a partially automated procedure tested on earlier data releases \citep[]{prochaska2004, prochaska2005}. Their cleanest ('statistical') sample contains 738 DLAs. We have chosen not to include objects that are listed in their non-statistical sample, as these can be towards QSOs with strong intrinsic absorption, or have z$_{abs} \sim${} z$_{QSO}$. Broad absorption line QSO (BALs) are known to be more strongly reddened in the rest-frame UV than non-BAL QSOs, and hence this list of DLAs is then cross-correlated with the BAL catalogue of \citet[]{gibson2009}, in order to avoid such BALs entering our analysis. Furthermore, we have restricted ourselves to those sightlines towards QSOs in the \citet[]{prochaska2009}{} sample that only exhibit one single DLA. Thus, we exclude the 59 respectively 5 cases, where two or three DLAs are found along the same line of sight. This leaves us with a sample of 526 SDSS DR5 QSO sightlines with securely identified DLA well suited for the analysis of their dust content.\\

\citet[]{noterdaeme2009}{} have surveyed DR7, and present a list of 1426 strong HI absorbers at redshifts 2.15$\leq${} z $\leq${} 5.2, of which 937 systems have log N(HI $\geq$ 20.3. Applying the same criteria for filtering out potentially problematic sightlines as above, we arrive at 731 SDSS DR7 QSO sightlines 
for this DLA sample.\footnote{The BAL catalogue can only exclude objects that were already in DR5. \citet[]{noterdaeme2009}{} make an effort to clean their samples of such sightlines, but we nonetheless found 44 entries with a match in the \citet[]{gibson2009}{} catalogue. Hence, we also visually inspected our
final sample of new DR7 sightlines to exclude obvious BALs, but found none.}  

While both lists have a substantial overlap (371 of Prochaska's sources appear
also in the Noterdaeme list, albeit 31 of these are estimated there to fall below the log N(HI $\geq$ 20.3 criterion), the two methods for 
finding DLAs differ significantly enough to warrant keeping both datasets separate. It should, however, be kept in
mind that the two samples are not independent from each other. Note also that the SDSS treatment of spectroscopic calibration has improved from DR5 to DR7, especially the spectrophotometric algorithms. Therefore, we have decided to strictly use only the versions of the spetra that have been processed by the DR7 pipeline.

\begin{figure*}
\includegraphics[angle=270,width=168mm]{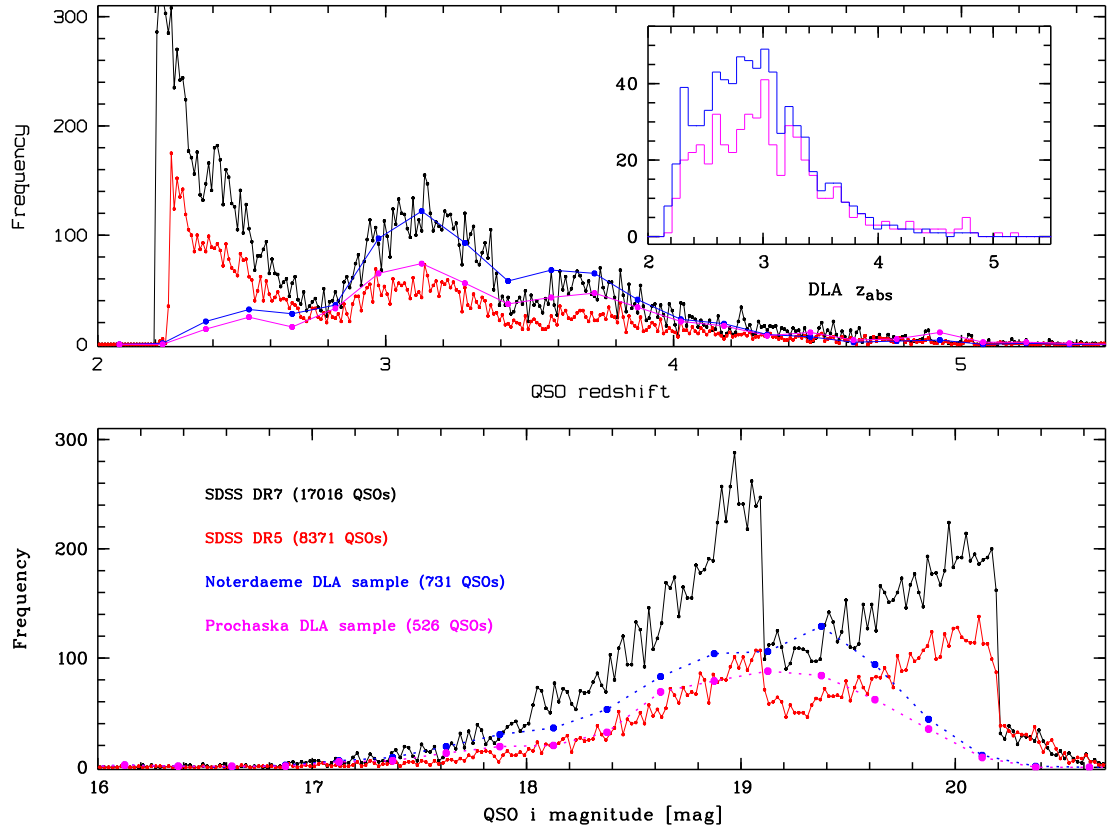}\caption{The distributions of the two DLA sample sightlines in QSO emission redshift and the $i${} band magnitudes. Plotted
are also the distributions for the underlying SDSS DR5 and DR7 QSOs. Note the factor
2 increase of sources from DR5 to DR7. The decrease of DLA containing sightlines towards lower redshifts, despite an increasing number of QSOs, is due to the ever shorter pathlengths over which HI absorption can be identified for the spectroscopic coverage of the instrument
trailing off at 3800 \AA{} on the blue side. In these, and the following figures, we have corrected the observed magnitudes for Galactic reddening. The insert shows the histogram of the absorbers in both samples. The majority of DLAs (80\%) are located between 2.3 $\leq${} z $\leq$ 3.3. The shape of this distribution directly affects the length of the leverage arm in the DLA restframes that is at our disposal in the composite spectra.}\label{imag_and_zem}
\end{figure*}

Figure \ref{imag_and_zem}{} shows the distribution of both samples in QSO emission redshift and the $i${} band magnitudes. In addition, we show the underlying SDSS QSO lists (cleaned of BALs and the DLAs from both lists, as detailed below) from which we can draw our matching spectra. The insert shows the histogram of the absorber redshifts for the two samples. The vast majority ($> 75\%$) of the DLAs are in the redshift range $\sim$2.2 $\leq$ z$_{abs} \leq$ 3.3.\\


\section{Procedure for determining extinction}\label{extinction_determination}
There are a variety of different methods to derive estimates for the reddening of each QSO spectrum, or the average reddening of the whole sample (e.g. \citet[]{pei1991, ellison2005, richards2003, murphy2004, vladilo2008}. We follow closely the suggestion of \citet[]{york2006}{} and \citet[]{vandenberk2008}, who have analysed the dust reddening in MgII absorbers towards SDSS QSOs. The basic principle is to define a sample of QSOs, where for each DLA quasar in question a close match in both emission redshift and $i$-band magnitude (and hence absolute i-band magnitude) is found, that does not contain a DLA. As \citet[]{york2006}{} point out, this matched non-absorber QSO spectrum will be of very similar data quality and hence sensitivity to similar levels of absorption at each wavelength, because of the regulation of exposure levels directly at the telescope and the exposure level check of all SDSS spectra before they are admitted to the database. \\
Given that there is a dispersion in the spectral slopes of QSO of similar magnitude and redshift, and hence a difficulty in matching spectra one by one, the strength of the method proposed here relies upon averaging over an as large as possible set of QSOs with individually different slopes to obtain a robust measure of the (geometric) {\it mean}  spectra for both the samples of QSOs containing DLAs and those free of DLAs. SDSS provides for both the largest homogeneous samples of DLAs as well as the largest reservoir of QSOs from which to draw comparison spectra.

\subsection{Constructing the non-absorber sample}
To find an appropriate match for a DLA-containing QSO spectrum, we search the SDSS DR5 database\footnote{Currently, the cataloque of \citet[]{gibson2009} only allows us to derive BAL-free samples up to DR5. Hence, we have to restrict ourselves to finding matches in DR5 only, as a reliable BAL search is beyond the scope of this paper. We have, in fact, checked for the effects of possible BAL inclusion by comparing samples drawn from DR5 strictly to those drawn including 'pure' DR7 QSOs, and concluded that indeed they lead to substantially 'redder' matching composites. Note that both the DR5 and DR7 DLA samples rely upon a BAL removal \citep[]{prochaska2009, noterdaeme2009}, and hence it is mandatory to implement such a check for BALs into the matching strategy, as well.  } after
cleaning it from BALs and also the DLAs found in both lists. This resulted in a pool of 8371 QSOs with redshifts 2.24$<$z$_{em} <$5.31, from which to draw the comparison spectra. We minimise the difference of the match to the QSO in the two-dimensional parameter space of the $i$ magnitude\footnote{We have used the values for the psf 'target' magnitudes and Galactic dust reddening as well as the QSO redshift SpecZ listed in the QSOConcordanceAll DR5 and DR7 datatables.} and emission redshift, defined as $D^{2} = (\frac{\Delta z_{em} / z_{em}}{0.007})^{2} + (\frac{\Delta i}{0.08})^{2}$. Note that we have corrected the $i${} magnitudes in each case for Galactic reddening. The normalisation factors in the distance metric were chosen such that the number of outliers
 in both quantities are minimal, as detailed in the following paragraph.\\

\begin{figure*}
\includegraphics[angle=270,width=168mm]{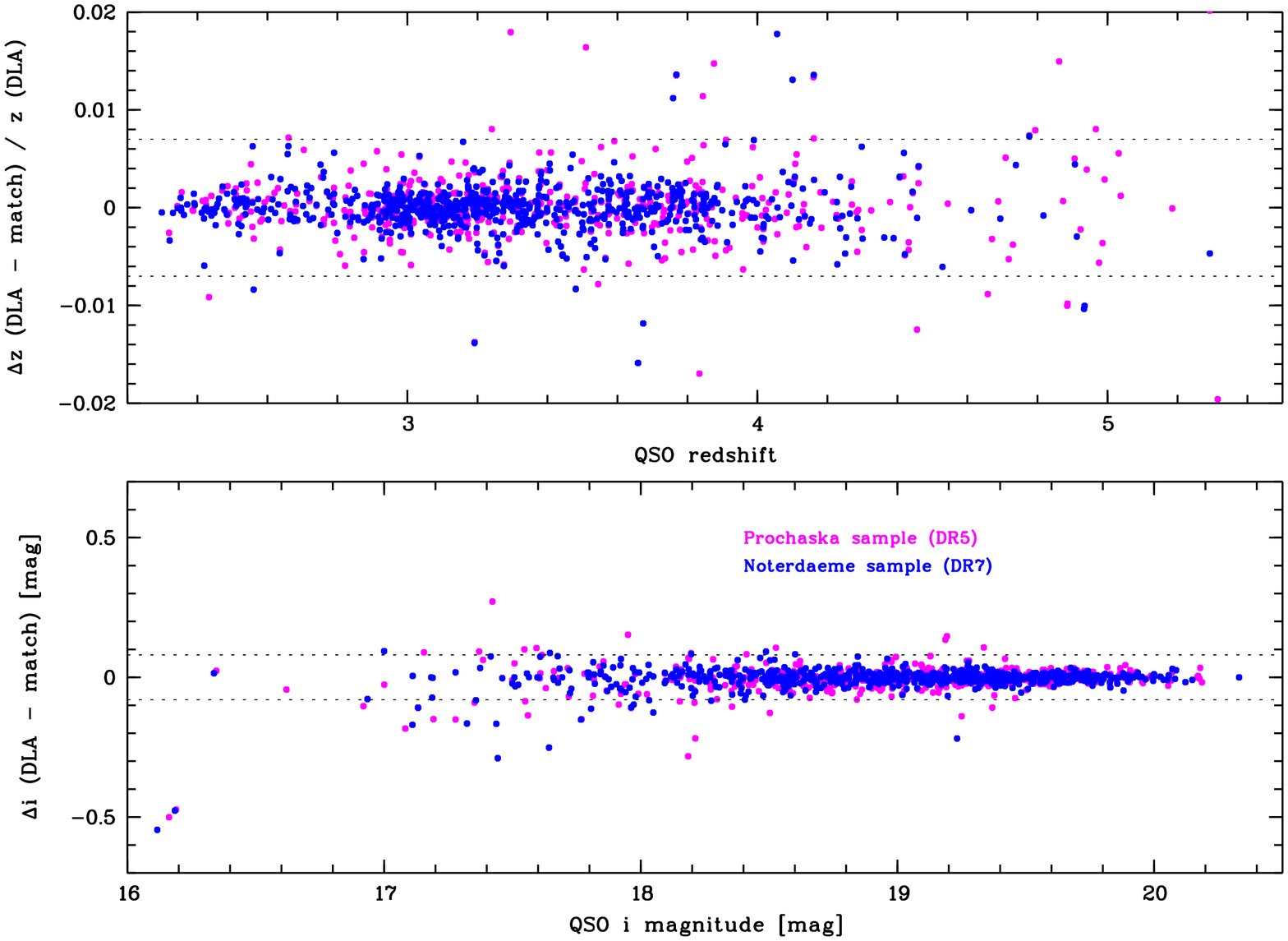}\caption{The difference in $i${} band magnitude and relative $z_{em}${} emission redshifts of the QSOs with DLAs and their matches for the DR5 (cyan) and DR7 (blue) samples. Note that there is a bias to pick fainter sources as matches, which
is especially pronounced towards the brighter objects ($i${} mag $\leq$18.0), where the number of QSOs rapidly dwindles (cf. figure \ref{imag_and_zem}). The dotted lines indicate the location of the boundaries that we deem acceptable for a good match.}\label{delta_z_delta_i} 
\end{figure*}

Figure \ref{delta_z_delta_i}{} shows the distribution of deviations between
the DLA containing spectra and their best matches in redshift and brightness for the two samples. 
The shape of the brightness distribution 
of the quasars in SDSS within the redshift range probed here, introduces 
a systematic Malmquist bias : in general there is a slightly higher chance 
to select a {\it fainter}{} match at {\it higher}{} redshift. This effect is particularly pronounced in
the brightest DLA QSOs, because the number of potential counterparts below 
$i${} mag $\leq$18.0 becomes too low to select an adequate match. Hence, we decided to exclude all DLA sightlines from the following analyses that deviate more than 0.08 in $i${} mag and/or 0.007 in relative redshift difference 
(dotted lines in the two panels of \ref{delta_z_delta_i}). This reduces the DR5 sample to 475, and the DR7 sample to 676 DLAs. The mean $\i${} band magnitude difference $\Delta i$(DLA - match) for these is -0.0029, and -0.0022 mag, whereas the mean relative redshift differences are $\Delta z$(QSO - match)/z = $9.3 \times 10^{-5}$, and $5.3 \times 10^{-5}$.\\

In principle, we could increase the number of matches for each DLA spectrum to more than one counterpart. However, the gain in S/N (or reduced scatter) for the ratio of DLA/match, dominated then by the DLA composite's noise, which cannot be reduced, does not compensate for the loss in matching precision : the more matches are being culled from the pool, the more severe the bias becomes. Of special importance here are they increasing redshift differences that may distort the precision at which the composite match spectrum reproduces the shape of the Hence, we have decided to utilise only the best match in each case.\\      

\subsection{Constructing composite spectra}\label{sample_construction}
The redshift distribution for the DLA systems in conjunction with the SDSS spectral coverage (3800 \AA{} $\leq \lambda \leq \sim 9500$\AA) allows in principle the construction of a composite spectrum in the DLAs' restframe from $\sim${} 600 \AA{} to $\sim${} 3000 \AA. Obviously, however, the flux distribution of each QSO blueward of 1215 \AA{} in the quasar restframe is strongly suppressed due to the neutral H column densities sampled inside the forest, in a manner specific to each sightline differently. Hence, in order to reliably determine the effects of dust reddening by comparison with an unreddened continuum, we have to restrict ourselves in each spectrum to the part redward of the beginning of the Lyman forest. We have therefore chosen to exclude all pixels in a spectrum that lie below 1240 \AA{} in the QSO restframe. This introduces a cutoff in the restframe of the absorbers that varies from spectrum to spectrum individually depending on the combination of z$_{em}${} and z$_{abs}$.\\
Geometric mean spectra were then generated from the absorber and non-absorber sample in the following manner : we extend the SDSS wavelength grid with a stepsize of 0.001 in log $\lambda${} to the lowest wavelengths covered by the spectra in the DLA restframe, and, as specified above, redward of the Lyman forest of the QSO itself. For each wavelength data point $\lambda _{DLA, RF}${} in this grid we compute for each sightline separately which pixel in the original spectrum comes to lie closest when transforming  by division through the absorber redshift, i.e. we take the nearest integer value of 
\begin{equation}\label{pixel_number}
N =  ( log{}  (\lambda _{DLA, RF} (1.0+z_{DLA}) - log{}  \lambda _{start} ) / 0.0001
\end{equation} 
where $\lambda _{start}$ denotes the wavelength of the first pixel in each spectrum, $z_{DLA}${} is the absorption redshift of the DLA, and $N${} is the pixel number starting from the blue end of the spectrum. Before computing the geometric mean of the flux distribution in the sample of the thus determined pixels, we clear it of those individual pixels flagged by the spectroscopic pipeline as possibly corrupted \citep[]{stoughton2002}. We have made no attempt in masking additional pixels, e.g. at the location of known absorption at different redshifts than the DLA. There is also no weighing of pixels according to e.g. S/N, as this possibly distorts the power-law nature of the combined spectrum. We have corrected the individual spectra for Galactic reddening \citep[]{schlegel1998}. The geometric mean composite spectra for the non-absorber sets were generated in exactly the same way, using the redshift of the corresponding DLA in each case for equation \ref{pixel_number}. Note that we use in each case  the spectra delivered by the DR7 pipeline, which includes the latest version of the spectrophotometric calibration. The advantage of the {\it geometric}{} mean is that it yields a meaningful average of the extinction at each wavelength by essentially averaging optical depths.\\

Note that our method does not rely upon any interpolation of the original spectra - at the price of sacrificing spectral resolution, because we bin together flux values that may reside in the DLA restframe separated by 1/2 of a pixel at the extreme. We have tested the effects of this by comparison of the resulting composite spectra to two cases of a more refined treatment :  I. rebinning each spectrum after deredshifting it, and II. computing the  weighted average of the flux values of the two pixels in the original spectra that always straddle one wavelength element in the new DLA restframe grid. As expected, the differences over large wavelength ranges are negligible, and manifest themselves to a noticable degree only on a two to three pixel scale. Hence, we believe our method, being as close as possible to the data, is robust enough not to have to rely on the intricacies of interpolation as in case I. Because we have to throw out pixels with fluxes f$_{\lambda} \leq${} 0.0, method II from above results in having to flag out twice as many datapoints, which is therefore deemed less acceptable despite the higher spectral resolution.\\  

\begin{figure*}
\includegraphics[angle=270,width=168mm]{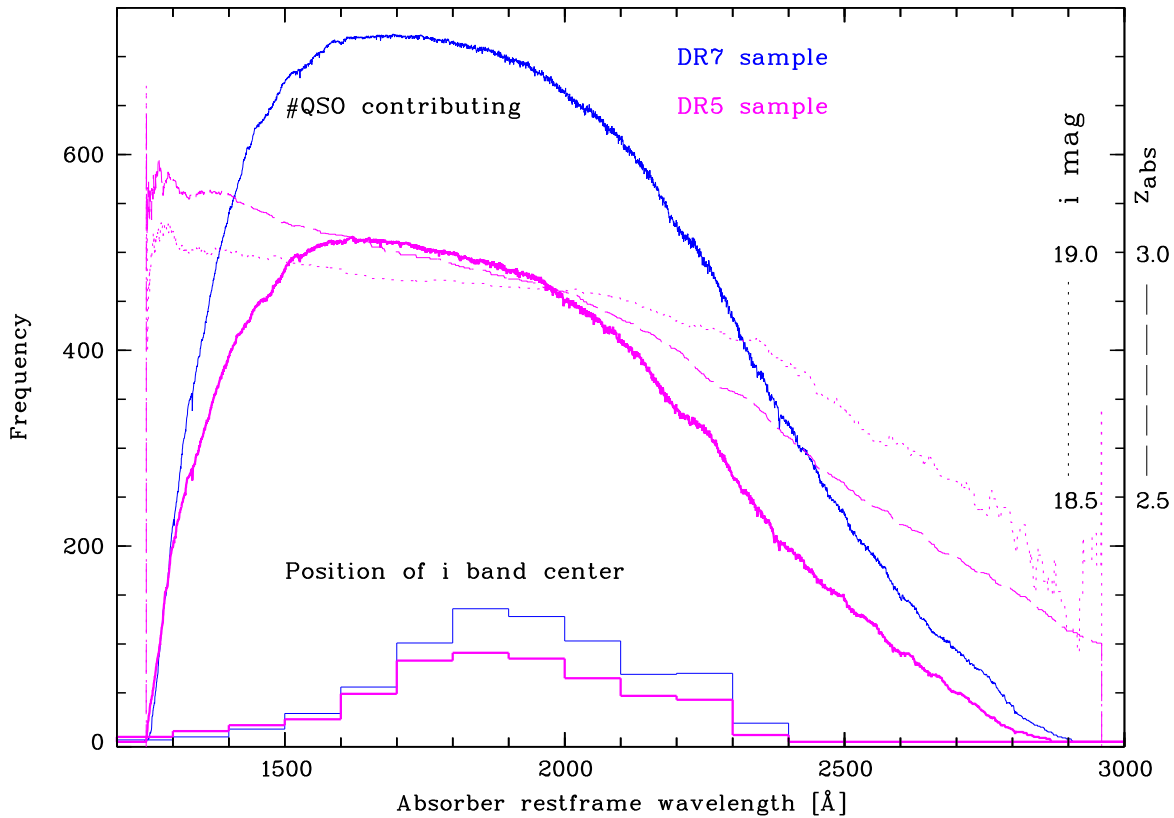}\caption{The number of spectra contributing to each wavelength bin for the construction of the composite spectra in the DLA
restframe (solid, thick lines). Also shown is the histogram of the position
of the central wavelength of the $i${} band (7481 \AA), that is used to find
 matches (frequency scale units for the histogram on the right). The dotted line marks the average brightness of the 
QSO sample going into the composite at each position (see scale on right axis), and hence explains the
rise towards long wavelength. This is accompanied by a similar drop in the average absorber redshift (dashed lines).}\label{number_contributing}
\end{figure*}

Figure \ref{number_contributing}{} shows the number of spectra contributing to each wavelength bin for the 
construction of the composite spectra in the DLA restframe. It also highlights to which location in these 
spectra the observed central wavelength of the $i${} band (7481 \AA) is shifted in the DLA frames. Furthermore we have indicated the average brightness of the QSOs contributing at each wavelength, and the 
average DLA absorber redshift. Note that the region covered by the bulk of the spectra extends from $\sim$1400 to $\sim$2300 \AA, and exhibits fairly constant values of the mean brightness and absorber redshifts. The need for staying redward of the Lyman $\alpha${} emission, and the decreasing high (restframe) wavelength with increasing redshift of each QSO, result of course in 
fewer spectra contributing to the composite towards the edges of the wavelength regime
covered by our samples. Hence, these spectral areas carry less weight due to their increased 
 noise/scatter in the reddening analyses (as detailed in 3.3.).\\
 
\subsection{Derivation of extinction}

\begin{figure*}
\includegraphics[angle=270,width=168mm]{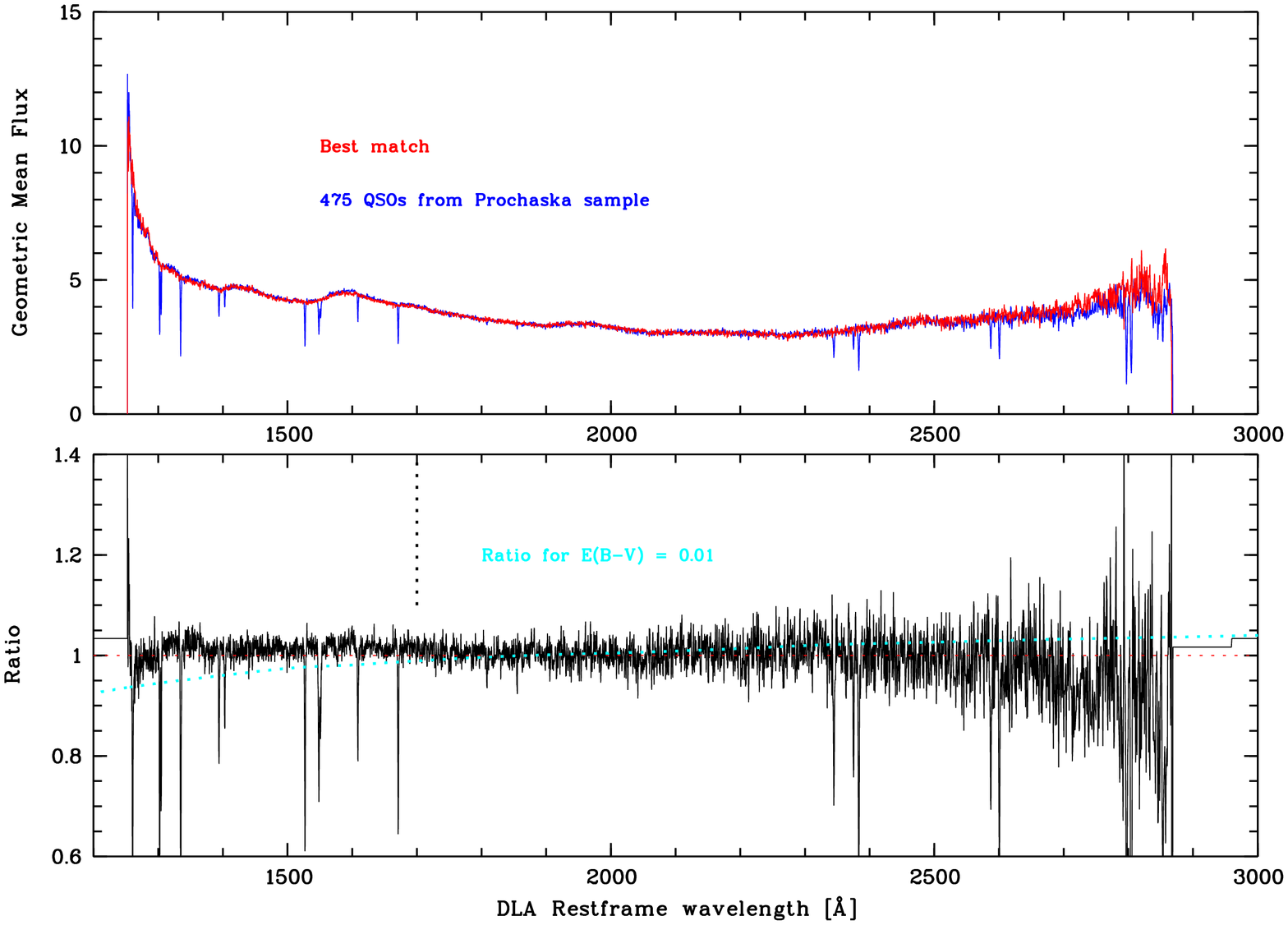}\caption{{\bf Upper panel:} The composite restframe spectra of the DLA DR5 sample (blue) and the corresponding sample of best matches (red). Before averaging the spectra, as detailed in the text, a Galactic dust extinction correction is applied. The stretched out broad emission lines have very similar shapes, indicating the suitability of the match for the analysis at hand. Only the DLA composite exhibits narrow metal line absorption features. The rise of the curves towards the red end is purely an effect of the larger average brightness of the sources at lower redshift, which are the sole contributors at this wavelength regime (cf. Figure \ref{imag_and_zem}). {\bf Lower panel:} Ratio of the the two composite spectra from the upper panel. The cyan dotted line indicates the ratio expected for an SMC extinction with E(B-V)=0.01 \citep[]{pei1992}, which exhibits clearly much more curvature than the data allow. The black dotted vertical line at 1700 \AA{} indicates the position bluewards of which the strong emission lines of CIV, possibly with contributions from Fe lines, and SiIV render the exact matching of spectral shape increasingly difficult (see text for details).}\label{prochaska_spectrum}
\end{figure*}

\begin{figure*}
\includegraphics[angle=270,width=168mm]{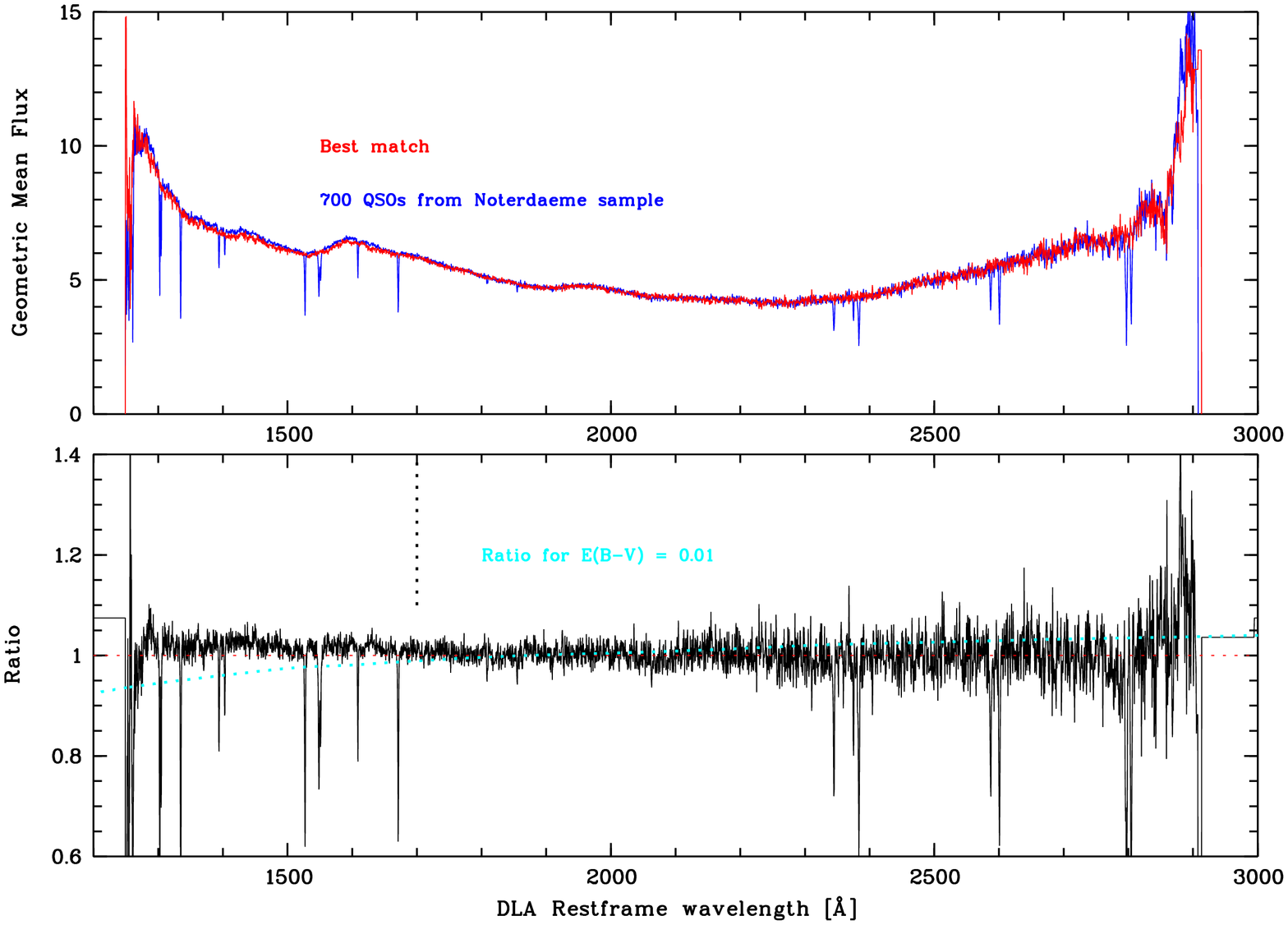}\caption{Same as figure \ref{prochaska_spectrum}{} for the DR7 sample.}\label{noterdaeme_spectrum}
\end{figure*}

The geometric mean composite spectra of the 475 (DR5) and 676 (DR7) absorber spectra, and the corresponding non-absorbers are shown in the top panel of Figures \ref{prochaska_spectrum}{} and \ref{noterdaeme_spectrum}. Note how the QSOs' emission lines for both samples are equally smeared out due to the different shifts applied for each specific combination of $z_{em}${} and $z_{abs}$, whereas only for the DLA sample the narrow absorption lines appear in the expected places. Because of the lower numbers of spectra contributing to each wavelength bin towards the blue and red end of the composite, the noise (or scatter) starts to increase. The rise of the mean flux towards the red side is an effect of the QSOs contributing to this part of the spectra being on average of lower redshift and brighter (cf. Figure \ref{imag_and_zem}). Additionally, the abrupt change in the flux beyond 2800 \AA{} can be attributed to the stack becoming essentially the average of a very few spectra from this point on.\\

The lower panels of Figures \ref{prochaska_spectrum}{} and \ref{noterdaeme_spectrum}{} show the ratios of the two composite spectra. The cyan dotted line indicates the ratio expected assuming an extinction of $<$E(B-V)$>${}  = 0.01 for an SMC curve \citep[]{pei1992}, and represents the upper limit on $<$E(B-V)$>${}  derived by \citet[]{murphy2004}. It is immediately obvious that very little room is left for the signature of {\it any}{} dust extinction in the DLA sample. We can quantitatively assess how much extinction can be tolerated by our data with the following procedure.\\

Suppose an unknown DLA stacked spectrum, $f_{DLA, nr} (\lambda)$ is reddened by dust within the absorbers, $f_{DLA, r}(\lambda)$. Then, the ratio of these two spectra would simply yield the extinction : $r_{0}(\lambda)=f_{DLA, nr}(\lambda) / f_{DLA, r} (\lambda) = 10^{0.4 A_{\lambda}}$. The underlying assumption for our method is that the stacked spectrum of the matches is a good representation of the unreddened DLA stack. Our matching scheme, however, leads us to an unreddened spectrum $f_{match, nr}(\lambda)$, that, by definition, has the same flux as the (supposedly) dust reddened DLA spectrum at some fiducial wavelength $\lambda_{0}$. Hence, the ratio of the two composites in this case is modified by the extinction $A_{\lambda_{0}}$ at this matching wavelength 
\begin{eqnarray}
r_{1}(\lambda)&=& f_{match, nr}(\lambda) / f_{DLA, r}(\lambda) \nonumber \\
             &=& 10^{0.4 \times (A_{\lambda} - A_{\lambda_{0}})}  
\end{eqnarray}

      
Given an extinction law, the ratio becomes a function of E(B-V) only : $r_{1}(\lambda) = r_{1}(E(B-V)$. Here, we have chosen the SMC extinction curve of \citet[]{pei1992}, and assume an $R_{V} = 2.93$ as typical for the total-to-selective ratio. Note that the choice of the SMC reddening curve is the most constraining regarding the absence of dust signatures - LMC or even MW dust type induces stronger reddening. Because we are fitting for the slope of this function, the potential offsets in $A_{\lambda_{0}}${} are of no relevance in this circumstance, and hence the normalisation of the ratio curve is not important.\\

\begin{figure*}
\includegraphics[angle=270,width=168mm]{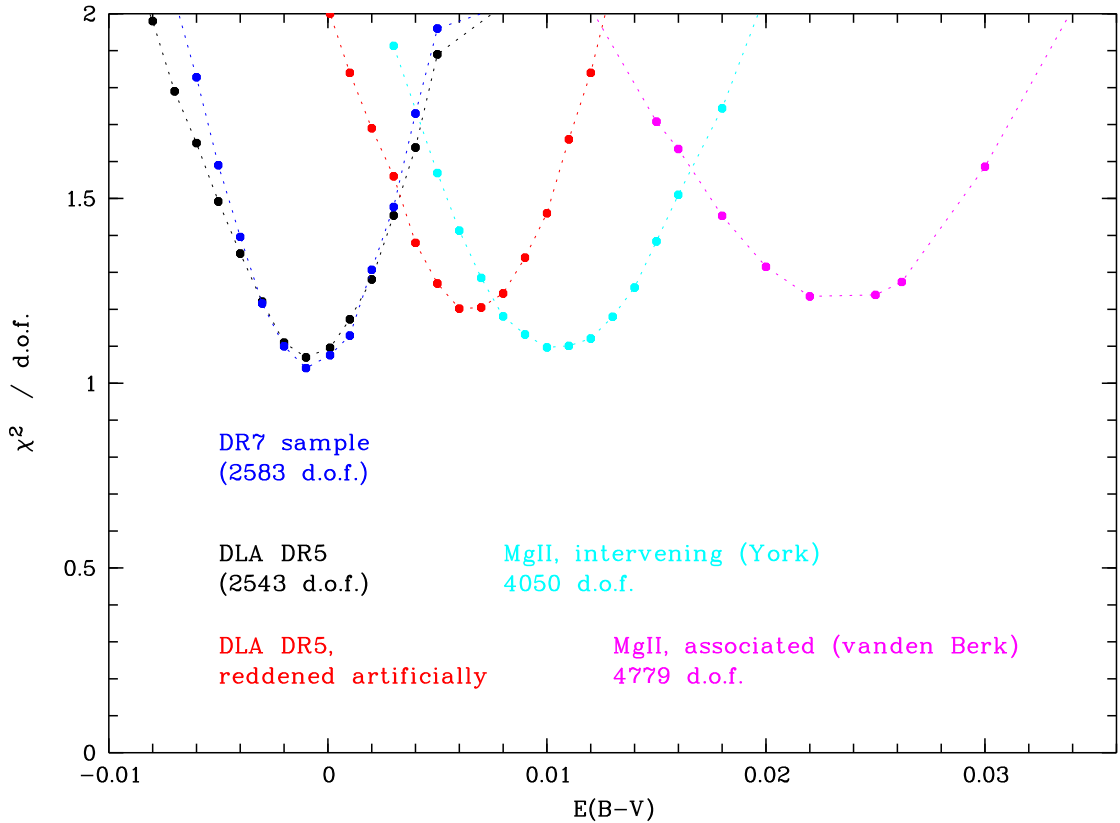}\caption{Results of the one-parameter fit for the dust reddening in various absorber types. Plotted are the $\chi ^2$ per degree of freedom distributions obtained by fitting the ratio of the composite spectra with a one-parameter extinction curve (for details see text). The two curves in magenta and cyan show the results for the samples of intervening and associated MgII absorbers in SDSS spectra \citep[]{york2005, vandenberk2008}. We recover the extinction $<$E(B-V)$>${}  measured by these authors. The black and blue curves are a result of the fit to the DLA sample. Note that we can clearly rule out any signs of extinction $<$E(B-V)$>${} $>${} 0.002 with very high confidence. The curve in red shows the result for the DR5 sample that has been artificially reddened by a lognormal distribution with mean $<$E(B-V)$>${} =6.3x10$^-3$, which is well recovered.}\label{chi2} 
\end{figure*}

Figure \ref{chi2}{} shows the results of these one-parameter fits for DLAs (this work) and two SDSS samples of MgII absorbers \citep[]{york2005, vandenberk2008}. To obtain the latter two, we have applied exactly the same averaging scheme as outlined above. Before the calculation of the goodness of fit statistic, $\chi^{2} / \nu${} (where $\nu${} denotes the degrees of freedom for the fit), we have taken out areas of $\pm$5 \AA{} from the known, narrow metal lines in the spectra. We find that we can constrain the amounts of extinction for those two samples of MgII absorbers very tightly : $<$E(B-V)$>${}  = 0.012$\pm$0.002 for the intervening, and $<$E(B-V)$>${}  = 0.025$\pm$0.003  for the associated cases. We therefore reproduce the results of \citet[]{york2005}{} and \citet[]{vandenberk2008}{}  well, despite the fact that we do not explicitly also remove regions of the spectra that are affected by strong absorption at redshifts different from the MgII absorbers. We take this test as a sign that both our stacking algorithms and the method to estimate the reddening are robust and accurate for these MgII samples. We note, however, that the lower redshift of these absorbers (compared to our samples of DLAs) may reduce the problem of the foreground reddening due to  the lower redshift path sampled, and the intrinsically lower density of absorbers in the low-z universe.\\

We have also applied this spectral method to the samples of DLAs and 'pool' QSOs used by Vladilo et al. (2008) 
for their photometric analysis. These authors have taken care to take out QSOs that may be afflicted by absorption unrelated to the DLAs. Again, we are able to reproduce their results very well, and derive an $<E(B-V)> = 6.2 \pm 1.8 \times 10^{- 3} $,
indicating that both methods yield similar results when applied to the same samples. Here, the strength of our method is specifically that all the unrelated effects will affect both the DLA and the pool spectra in the same way, and hence we choose to increase the sample size to ensure that these effects are averaged out. We are making use of both the
largest databases for DLAs {\it and}{} comparison spectra (increasing the number of DLAs by almost a factor of 3 for the DR7 sample, and
the number of the objects in the comparison pool by a factor of 5 compared to the Vladilo samples).\\

A word of caution is necessary when trying to quantify the confidence intervals for our results. Nominally, the number of degrees of freedom (d.o.f.) for the two samples are 2543 (DR7) and 2583 (DR7), after taking out the pixels affected by metal-line absorption, and restricting ourselves to the wavelength region below 2500 \AA{} to avoid the large drop in both numbers of spectra contributing and average QSO brightness. However, as we have detailed in the section on the construction of the composite spectra, adjacent pixels are not truly independent of each other. We have tested the robustness of the fits by binning the composites (with bin sizes up to 5 pixels, hence reducing the d.o.f. by the same factor). Hence, we arrive at measurements of the mean $<$E(B-V)$>${} =-0.0013$\pm$ 0.0025 (DR5) and $<$E(B-V)$>${} =-0.0017$\pm$0.0022 (DR7) for the full samples\footnote{Of course, negative reddening values are per se unphysical, but the natural outcome of the fitting routine. These are largely driven by the fact that the regions in the spectrum with the strong QSO emission lines between 1300 and 1700 \AA{} cannot be expected to be perfectly matching, given the complicated line emission structures of quasars. This difficulty, and not sample size, probably also sets the limiting precision of our method.}, where the error budget adopts a generous 3$\sigma${} level for the approach with the 5-pixel bin.\\

\begin{figure*}
\includegraphics[angle=270,width=168mm]{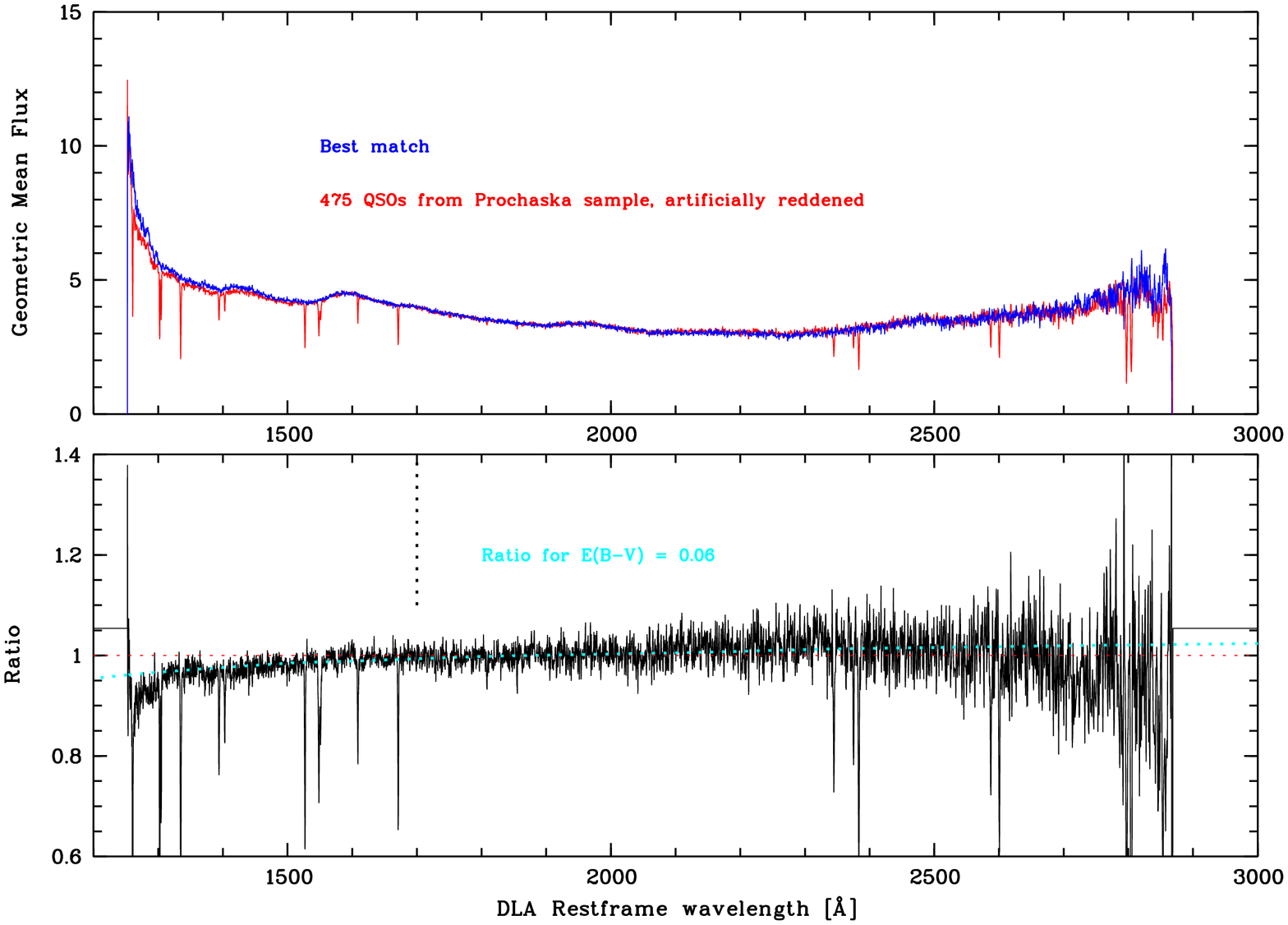}\caption{The result of artificially reddening the  DR5 DLA sample spectra with each QSO subjected to an $<$E(B-V)$>${}  drawn randomally from a lognormal distribution of mean log $<$E(B-V)$>${} =-2.2 and width 0.4 dex, resembling the values found by \citet[]{vladilo2008}{} and \citet[]{pontzen2009}. The reddening is not only clearly detected by our method, but also the input value is returned accurately. Labelling and underlying comparison spectrum just like in Figure \ref{prochaska_spectrum}.}\label{artificial_reddening}
\end{figure*}

In order to test whether we could have detected an $<$E(B-V)$>${}  distribution as predicted by \citet[]{pontzen2009}{} and measured by \citet[]{vladilo2008}, we have artificially reddened each DLA quasar spectrum by a value drawn from a lognormal distribution of width 0.4 dex and mean of log $<E(B-V)> = -2.2$, broadly consisted with those two studies. Figure \ref{artificial_reddening}{} shows that we not only could clearly detect such a reddening, but also could retrieve the input value very accurately (red curve in Figure \ref{chi2}). In fact, the two $\chi^2${} curves for our spectral analyses and the fake reddening are excluded by several $\sigma$. \\

We have also performed a 'null test' by comparing the composite spectra of the matched samples with composites created from the samples of the 2nd best, 3rd best, etc.... matching samples. As expected, the analysis of the ratios of these stacked spectra yields no detectable reddening, and the range of $<$E(B-V)$>${}  values obtained by this procedure ranges from -0.002 $\leq$$<$E(B-V)$>${} $\leq$0.002 for DR5, and -0.0018 $\leq$$<$E(B-V)$>${} $\leq$0.001 for DR7. Hence we encompass the two measurements with the DLA samples. We note in passing that the main difficulty in matching the  shape of the composites and their comparison counterparts lies in the strong emission lines of the QSOs, especially blueward of $\sim$1700 \AA, where the Fe-complexes have a profound influence on the flux above the continuum. Indeed, when inspecting the ratio curves for the two samples, around that wavelength there seems to appear a 'kink' in them, which may be caused by a mismatch in the emission lines.  This is the main driver towards these (very low) 'negative' $<$E(B-V)$>${}  values we obtain formally from the fitting. If we restrict ourselves to either the ratios blueward or redward of this feature, the fitting results in exact null values (at the expense of leverage in wavelength).\\

One possible complication in matching the quasars' luminosities based upon their emission redshifts and the observed i band magnitudes is that the latter could systematically underestimate the true continuum flux for the DLA sample compared to their non-DLA counterparts because of a higher fraction of strong metal absorption lines due to the presence of the DLA. A rough estimate of the size of this effect indicates that in sightlines where a very strong absorber (of the order of 5 \AA{} restframe equivalent width) is present in the i band, it will result in a photometric measurement that is decreased from the 'normal' QSO brightness by 2\%. Thus we may in such cases be selecting QSOs as matches that are systematically too faint. We have tested whether such a possible bias may influence our analysis of the dust reddening signal by creating samples of matches following exactly the procedures described above, but introducing an artificial offset $\delta i${} in the i band magnitude by minimising $\Delta (i_{DLA} + \delta i) - i_{match}$. Even shifts as large as $\delta i$=-0.2 magnitudes do not change the overall shape of the composite spectra, yielding almost exactly the same results for the reddening estimates when fitting for the slope of the ratio curves.


\section{Extinction analyses for subsamples}
While the full samples of DLA sightlines do not show any appreciable sign of dust reddening to a high degree of accuracy, it is still possible that there are certain subsets of DLAs and/or sightlines that do exhibit an amount of wavelength dependent extinction, which is simply too  small to be detected when averaging over such large datasets. In order to check for such potential trends, we have divided the DLA lists into subsamples according to various criteria concerning the DLAs themselves (namely absorption redshift, HI column density, and the ab- or presence of metal lines) or the underlying QSO sightlines (apparent brightness or emission redshift). To keep a high enough number of DLAs we have in almost all cases decided to simply split the full samples into two halves.\\

\begin{figure*}
\includegraphics[angle=270,width=168mm]{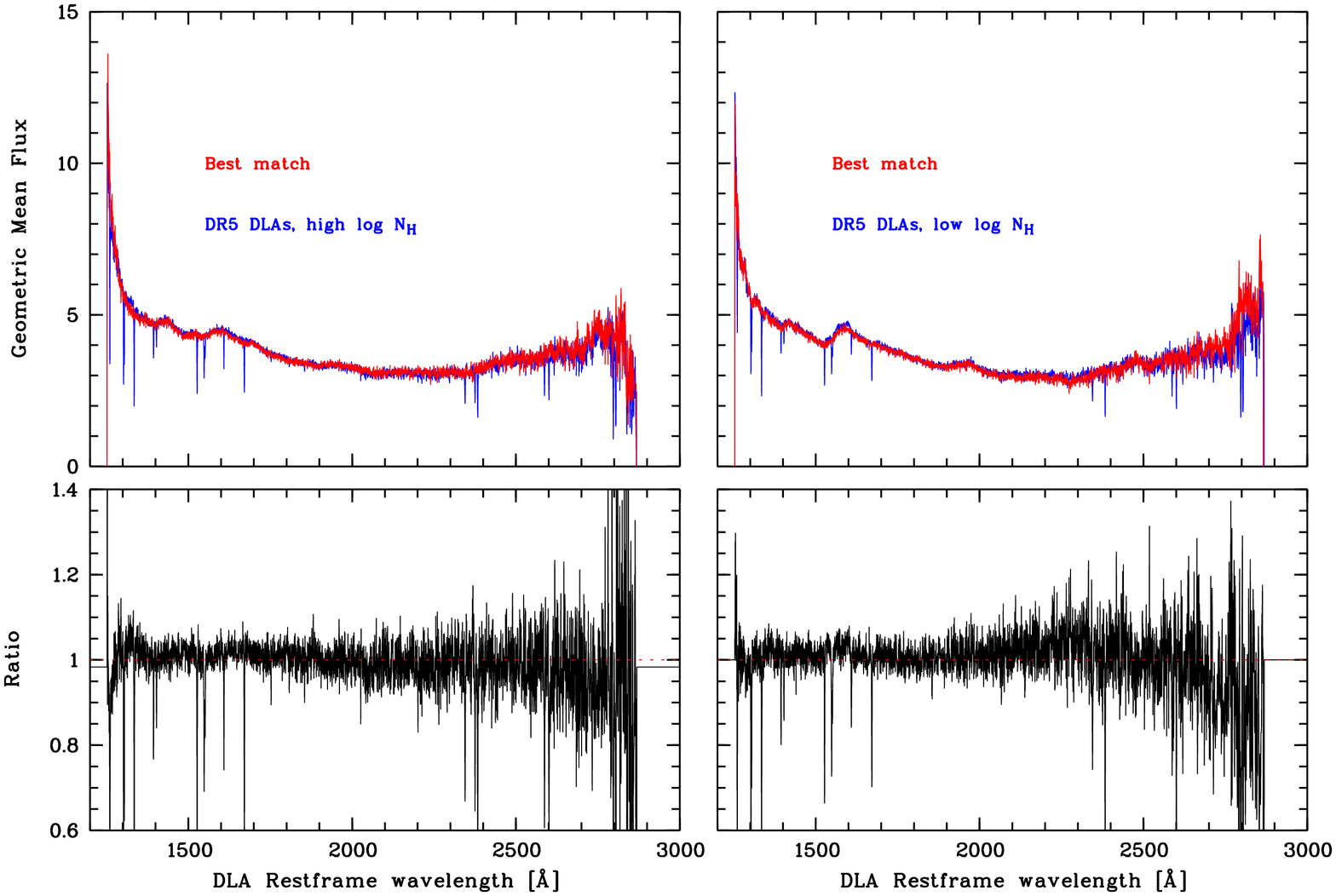}\caption{Comparison of the geometric mean spectra and their ratios for the subsamples with the {\bf highest} (left panels) and {\bf lowest} (right) column densities log N$_{HI}$. Note that for this, and the following plots (with the exception of the metal-no metal distinction), each of the subsamples simply contains half the number of sightlines as the original full sample.}\label{subsample_NH} 
\end{figure*}

Figure \ref{subsample_NH}{} shows the results of a split (for the DR5 sample)\footnote{We have performed the subsample analyses for both DR5 and DR7, yielding the same results, but only show the results for DR5 here as examples.}{} into the highest (left panel) and lowest (right panel) HI column densities for the DLA. The mean of these subsamples are at log N$_{HI}$=20.91, and log N$_{HI}$=20.42. Clearly, there is no difference in the fact that there is a null detection for both subsamples.\\

Figure \ref{subsample_zabs}{} shows how differentiating the samples according to the absorber redshifts affects the dust reddening analysis. Again, both for the low and the high redshift subsamples (mean z$_{abs}$=2.63, and 3.39, respectively) there is no reddening detectable. Note how the higher redshift sample leads to a lower wavelength cut-off, and due to the generally fainter spectra a slightly higher noise in the composite.\\

Figure \ref{subsample_metals}{} demonstrates that the DLAs classified by \citet[]{prochaska2009}{} as containing no visible narrow metal absorption lines in individual SDSS spectra, do also not exhibit any such signs in their  composite. Only 88 of the 475 DLAs fall into this class (18.5\%), and hence the noise in the composite is much higher than in the other subsamples. Yet, both classes do not show reddening.\\

The separation into QSOs of low and high redshifts (shown in Figure \ref{subsample_zem}, mean z$_{abs}$=2.31 and 3.78) is similar to the cut in QSO brightness (mean $i$ band psf magnitude of $i$=18.62, and 19.48 mag, corrected for Galactic reddening), because of the trend for fainter sources with higher redshift. The subsample with the brightest QSOs (for DR5)\footnote{The same cut in the DR7 sample does not show such a detection.} is the only one that does show a positive value of the colour excess, but not signifcantly different from a null detection : $<$E(B-V)$>${} =0.0012$\pm$0.0025.\\

To summarise, none of these simple subselections do give rise to the assumption that there are significant trends in reddening regarding either the observational constraints (brightness and  emission redshift of sources) or DLA properties (absorption redshift, additional narrow metal line absorption, and HI column density).

\begin{figure*}
\includegraphics[angle=270,width=168mm]{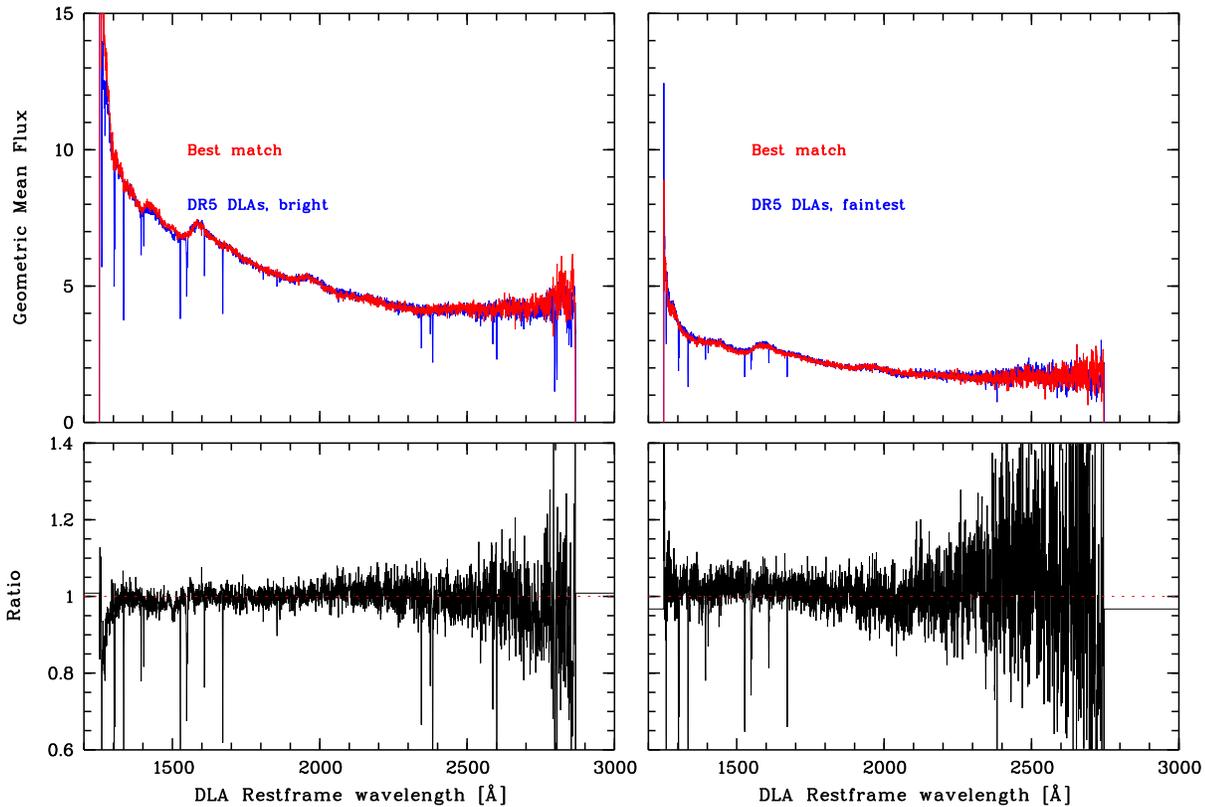}\caption{Comparison of the geometric mean spectra and their ratios for the subsamples with the {\bf brightest} (left panels) and {\bf faintest} (right) QSOs. Obviously, the ratio for the fainter subsample is noisier, yet no systematic of reddening in either subsample can be discerned.}\label{subsample_brightness}
\end{figure*}

\begin{figure*}
\includegraphics[angle=270,width=168mm]{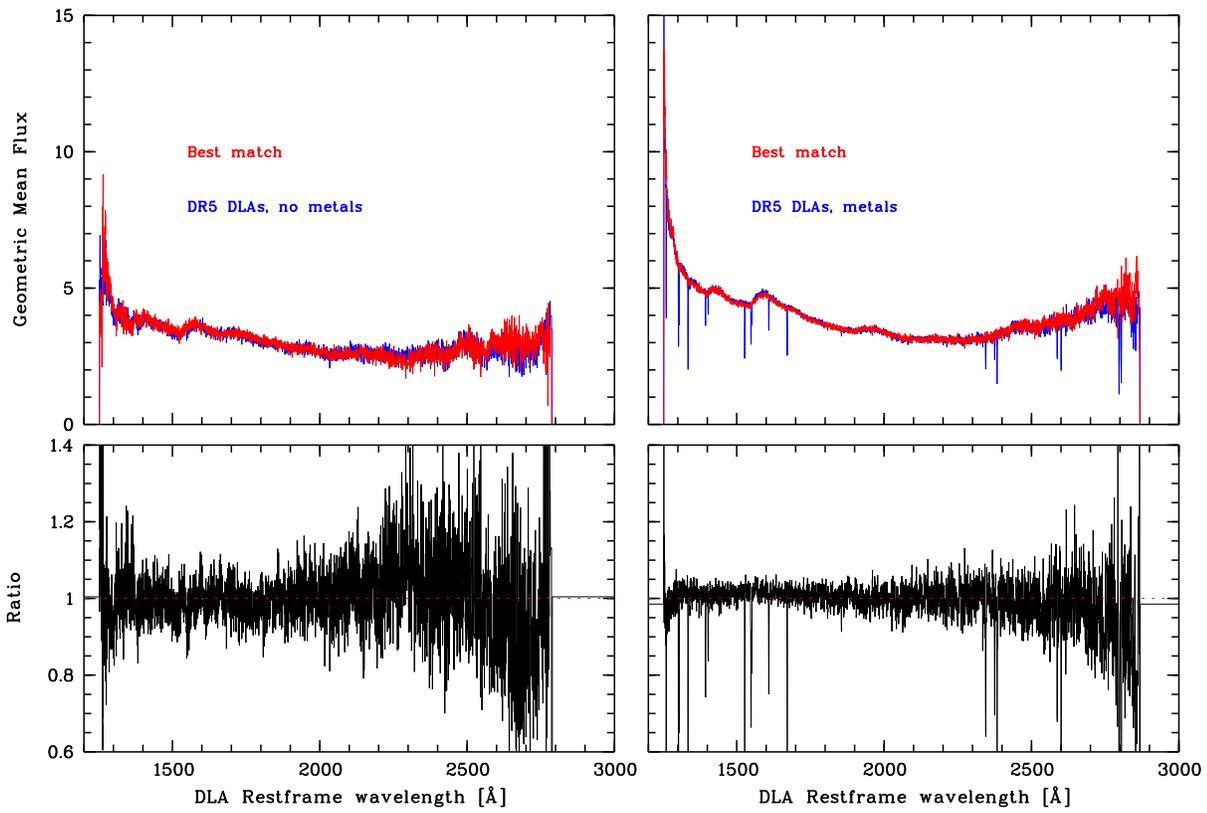}\caption{Comparison of the geometric mean spectra and their ratios for the subsamples with the DLA that do not exhibit sign of {\bf metal-line absorption} (left panels, 88 DLAs in total) and the ones that do (right panels). Note the conspicuous absence of {\it any}{} sign of metal-line absorption in the DLA stack on the left.}\label{subsample_metals}
\end{figure*}

\begin{figure*}
\includegraphics[angle=270,width=168mm]{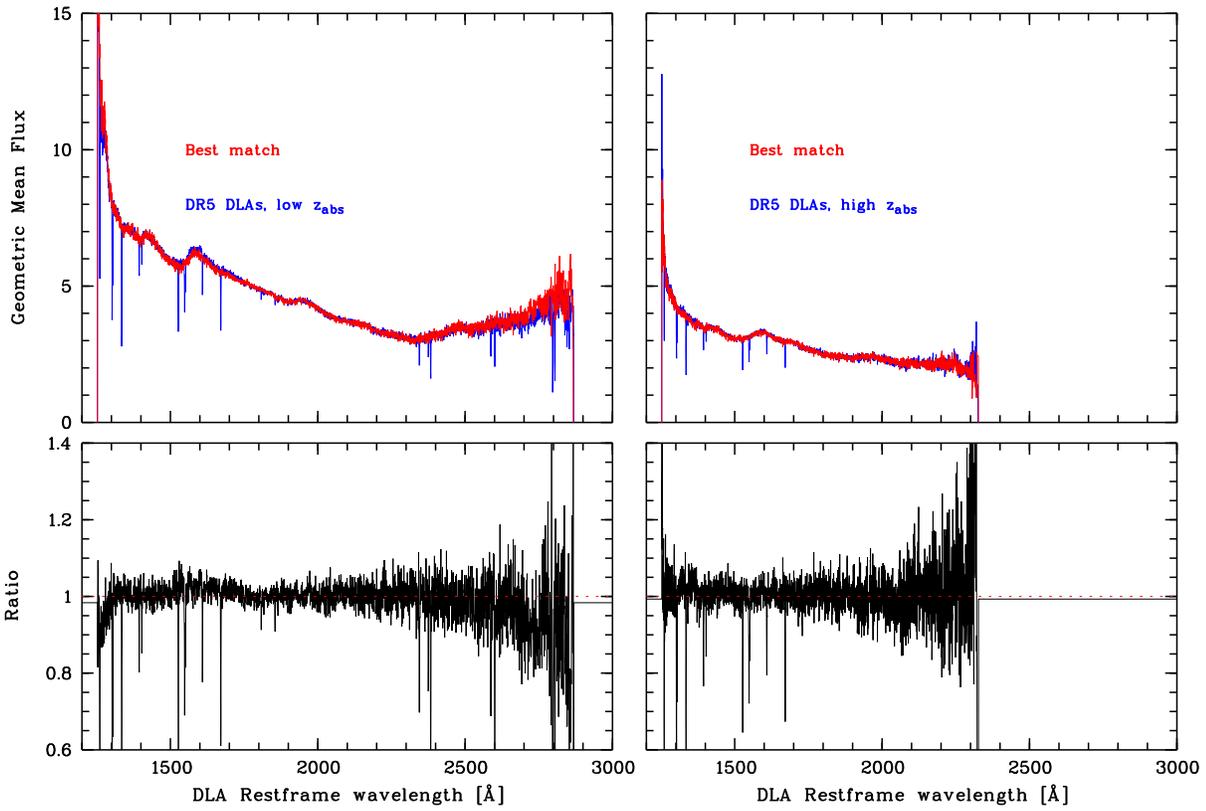}\caption{Comparison of the geometric mean spectra and their ratios for the subsamples with the {\bf lowest} (left panels) and {\bf highest} (right) DLA absorber redshifts. Notice the lower cut-off wavelength for in the DLA restframe for absorbers with higher redshifts. Despite the fact that hence the 'kink' beyond $\sim 2500 \AA${} in the stacked spectrum of the low redshift subsample gets even more pronounced than in the overall sample, the matched spectrum still follows this quite accurately.}\label{subsample_zabs}
\end{figure*}

\begin{figure*}
\includegraphics[angle=270,width=168mm]{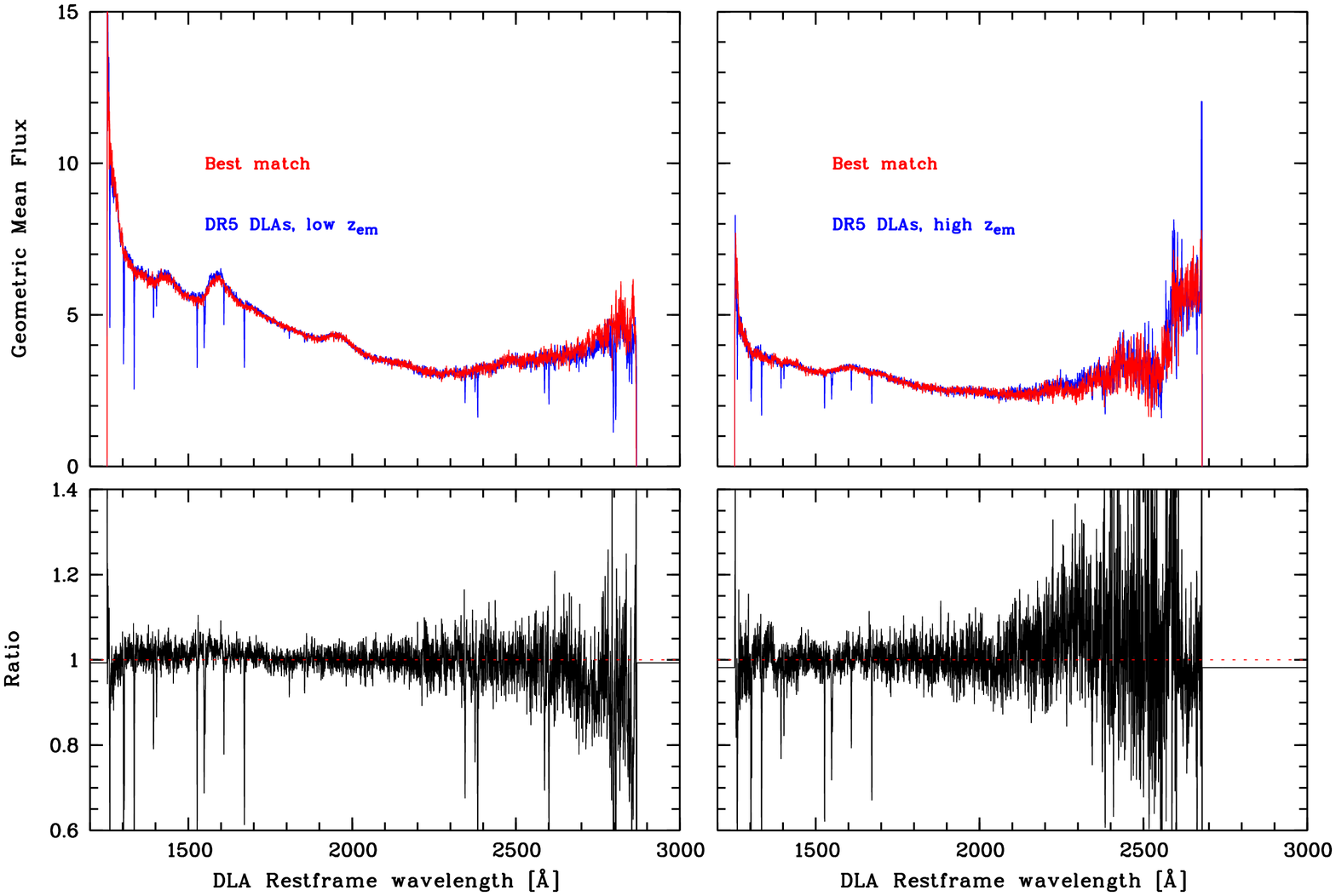}\caption{Comparison of the geometric mean spectra and their ratios for the subsamples with the {\bf lowest} (left panels) and {\bf highest} (right) QSO emission redshifts. Note the rapidly increasing noise for the high redshift subsample towards the red part of the composite. The separation into QSOs of low and high redshifts is similar to the cut in QSO brightness. The subsample with the brightest QSOs (for DR5) is the only one that does show a positive value of the colour excess, but not signifcantly different from a null detection : $<$E(B-V)$>${} =0.0012$\pm$0.0025.}\label{subsample_zem}
\end{figure*}

\section{Summary and conclusions}\label{summary}
While our analysis technique reproduces well past results for other classes of
absorbers at different redshifts \citep[]{vandenberk2008, york2006}, and DLAs using other methods \citep[]{murphy2004} and \citep[]{vladilo2008} where some extinction could be detected, we could no detect
any significant extinction in samples of DLAs drawn from the Data
Releases 5 and 7 of the SDSS. These findings are based on the largest DLA
sample ever studied, but also on the largest sample of quasars from
which to draw a reference spectrum. The implication is that most of
the DLAs observed in SDSS have a very small dust content.\\


\citet[]{york2006}{} have analysed SDSS spectra with intervening strong MgII absorbers, and detect clear dust reddening. More specifically, their subsamples with the highest MgII equivalent width (subsamples 8 and 25 in their notation, Table A4) show the highest degrees of reddening ($<$E(B-V)$>${} =0.032 and 0.085). These strong MgII absorbers are then classified, based on the column densities of the accompanying metal lines, as akin to low redshift DLAs. We note however, that a direct measurement of the HI absorption in these absorbers was not possible due to the low wavelength cutoff of SDSS, and the mean HI column densities for the subsamples are derived by using an SMC relationship between $<$E(B-V)$>${}  and HI. \citet[]{rao2006}{} also caution that not all high EW MgII absorbers are indeed DLAs, although \citet[]{york2006}{} point out that the specific requirement of the QSOs probed by \citet[]{rao2006}{} with HST being UV-bright introduces a bias against red QSOs, that may have lead them to an underestimate of the DLA fraction. Hence, it is a priori not clear whether our results contradict these findings. Also note that there may be a redshift evolution effect. In fact, \citet[]{menard2008}{} have noted that the rest-frame colour excess E(B-V) for their MgII samples shows a strong decline with increasing redshift (their figure 10, lower right panel). Thus, it is possible that the difference in the mean redshifts of our samples (z$_{abs} \sim$3.1) compared to the lower z$_{abs} \sim$1.4 of \citet[]{york2006}{} may partly explain the seemingly discrepant results. \\

Yet another strategy involving  spectral analysis is fitting the QSO continuum spectra by a power-law, and checking for deviations in the slope of these for spectra with and without DLAs. Employing this technique, \citet[]{murphy2004} can place an upper limit for a sample of 70 DLAs from SDSS DR2 of $<$E(B-V)$>${} =0.01. Note that we can rule out such a high value for our samples with a very high degree of confidence. \\

In addition, there have been claims of firm detections of dust reddening towards sightlines
afflicted by intervening DLAs. Specifically, \citet[]{vladilo2008}{} report a mean E(r-z)=27($\pm$9)x10$^{-3}${} for a subsample of the same DR5 DLA sample that we use. When applying our method to these subsamples of DLAs {\it and}{} the subsample of DR5 'pool' QSOs for the constructing of the comparison spectrum, we recover their results. Note that \citet[]{vladilo2008}{} have removed all quasars from their samples that may have additional absorption unrelated to the DLA, reducing the number of objects in each sample by a large fraction. We conclude that the differences do not reside in the methods, but the sample selection. We note that the reddening distribution of undesired interlopers probably has a tail with a small number of objects exhibiting much larger than the typical reddening expected from DLA systems. It may hence be that for such low frequency events, that cases of e.g. BAL QSOs, despite efforts to filter them out, enter the DLA samples more often than the comparison pool, and hence could wash out the signal. Increasing the sample size reduces the possible impact of such 'special', very red QSOs. 

Our result is also at odds with expectations from the evolution of high-redshift
galaxies (but see \citet[]{pontzen2008}. It should be emphasised however that there are two separate
issues: i) what is the dust content of the quasar absorbers we know of
today and ii) which fraction of the quasar absorbers are missed
because of the extinction they induce on their background quasar such that it is
not selected in the first place. Indeed, models of galactic chemical
evolution (e.g. \citet[]{prantzos2000} predict the existence of
regions of high metallicity and column density which are mostly not
detected in current DLA surveys \citep[]{boisse1998}. One way to
reconcile these two findings is to invoke a dust-bias in the
observations. These apparent discrepancies could indeed be explained
by the dust content of the absorbers themselves: they could absorb
enough light from the quasar to make the latter too faint to be
included in current quasar samples, which barely go beyond magnitude
g$\sim$20.2. In fact, an extension of the \citet[]{fall1993} calculations
taking into account recent observations addresses both these issues:
\citet[]{vladilo2005} propose that whilst the dust content of the
DLAs in current samples is not high, the missing fraction is possibly
quite important ranging from 30\% to 50\% at $z=1.8-3.0$. One way to
to test such a hypothesis is to look at fainter objects than what can be
probed with SDSS. Using a Bayesian analysis of these issues in light of 
different techniques and datasets \citet[]{pontzen2009} posit that 
probably less than 10\%{} of DLAs have been missed in current optical samples, and also derive an $<$E(B-V)$>${}  posterior probability distribution peaking strongly around log $<$E(B-V)$>${} $\sim -2.0$.\\  

On the other hand, if the absence of the detection of extinction in
DLAs is not due to an observational bias, then selecting in absorption
implies that we are picking a population of high-redshift galaxies which
have probably undergone little star formation, e.g. \citet[]{weatherley2005} and \citet[]{wolfe2006}.  This is difficult to
reconcile with the fact that these system are gas-rich by
construction.

\section{Acknowledgments}
This work has benefitted from support of the "Agence Nationale de la Recherche" with reference ANR-08-BLAN-0316-01. We would like to thank Bruno Milliard, Jean-Michel-Deharveng, Jochen Liske, and Michael Murphy for
helpful comments refining the analysis substantially. We thank the referee for a thorough review of the results 
and valuable suggestions, that improved the mansucript.\\

Funding for the SDSS and SDSS-II has been provided by the Alfred P. Sloan Foundation, the Participating Institutions, the National Science Foundation, the U.S. Department of Energy, the National Aeronautics and Space Administration, the Japanese Monbukagakusho, the Max Planck Society, and the Higher Education Funding Council for England. The SDSS Web Site is http://www.sdss.org/. \\

The SDSS is managed by the Astrophysical Research Consortium for the Participating Institutions. The Participating Institutions are the American Museum of Natural History, Astrophysical Institute Potsdam, University of Basel, University of Cambridge, Case Western Reserve University, University of Chicago, Drexel University, Fermilab, the Institute for Advanced Study, the Japan Participation Group, Johns Hopkins University, the Joint Institute for Nuclear Astrophysics, the Kavli Institute for Particle Astrophysics and Cosmology, the Korean Scientist Group, the Chinese Academy of Sciences (LAMOST), Los Alamos National Laboratory, the Max-Planck-Institute for Astronomy (MPIA), the Max-Planck-Institute for Astrophysics (MPA), New Mexico State University, Ohio State University, University of Pittsburgh, University of Portsmouth, Princeton University, the United States Naval Observatory, and the University of Washington.


\bsp

\label{lastpage}

\end{document}